\documentclass{article}
\usepackage{spconf,amsmath,amssymb,multirow,graphicx}
\usepackage{algorithm}
\usepackage{algpseudocode}
\usepackage{enumitem}
\setlist[itemize]{noitemsep}
\setlist[enumerate]{noitemsep}

\DeclareMathOperator*{\argmax}{arg\,max}

\title{Optimizing Byte-level Representation for End-to-end ASR}
\name{Roger Hsiao, Liuhui Deng, Erik McDermott, Ruchir Travadi, Xiaodan Zhuang}
\address{Apple}

\begin{document}
%
\maketitle
\begin{abstract}
  We propose a novel approach to optimizing a byte-level representation for end-to-end automatic speech recognition (ASR). Byte-level representation is often used by large scale multilingual ASR systems when the character set of the supported languages is large. The compactness and universality of byte-level representation allow the ASR models to use smaller output vocabularies and therefore, provide more flexibility. UTF-8 is a commonly used byte-level representation for multilingual ASR, but it is not designed to optimize machine learning tasks directly. By using auto-encoder and vector quantization, we show that we can optimize a byte-level representation for ASR and achieve better accuracy. Our proposed framework can incorporate information from different modalities, and provides an error correction mechanism. In an English/Mandarin dictation task, we show that a bilingual ASR model built with this approach can outperform UTF-8 representation by 5\% relative in error rate.
\end{abstract}

\section{Introduction}
\label{sect:intro}

End-to-end (E2E) neural networks are flexible and accurate models for multilingual automatic speech recognition (ASR). The output of such a multilingual model is often
unions of characters or subwords of the supported languages. However, as the number of languages increases, the size of the output layer increases, which can
negatively affect compute, memory usage and asset size. This problem is more prominent when the system supports languages that have large character sets, such as Chinese, Japanese and
Korean (CJK). To tackle this problem, previous work proposed the use of byte level representation for E2E ASR~\cite{li2019, deng2022}. By
using UTF-8~\cite{unicode2011} codewords as the underlying base tokens, the output vocabulary is no longer constrained by the character sets of each language, allowing developers to choose
a vocabulary size based on compute, and memory constraints. One well-known multilingual ASR system that uses UTF-8 subwords is Whisper~\cite{pmlr-v202-radford23a}.

UTF-8 aims to represent all the characters used in major languages. The encoding and decoding processes are designed to be simple and efficient. UTF-8 is a 
variable length prefix code where each character is represented by one to four bytes. Most byte sequences are not valid UTF-8 strings, and the UTF-8 decoder needs to detect invalid sequences. UTF-8 also provides backward compatibility, where ASCII characters are represented by a single byte and they are the same as the ASCII encoding.
While UTF-8 has proven to be an effective output representation for ASR, it is unclear whether it is optimal.
For example, characters with similar pronunciations or meaning are not guaranteed to share the same prefixes.
In addition, the large number of invalid byte sequences means the model needs to identify valid UTF-8 strings, an additional burden.

In this work, we explore the possibility of optimizing a byte level representation for E2E ASR with a data driven approach.
We believe an ideal byte level representation should:
\begin{enumerate}
  \item Be optimized for target recognition/classification task;
  \item Consider all available information;
  \item Provide error correction.
\end{enumerate}
For the first point, we think the representation should be optimized for the target machine learning task. The design should
boost accuracy, and be data driven. For the second point, we think the design should
consider all available information. For example, if the representation is designed for ASR, it should incorporate the information from both text and audio.
Ideally, the framework should be flexible enough to incorporate any available information. For ASR, it means it should be able to take
lexicon, phonemes or other useful information into account. For the last property, we believe the representation should provide an error correction mechanism
to handle the cases when the model produces an invalid sequence, and the recovery should aim for minimal error.

In this paper, we propose a novel representation learning approach with a vector quantized auto-encoder. We show that
we can optimize the byte level representation for ASR with available text and audio data. The framework is flexible enough so it can be extended to incorporate
side information, such as a lexicon. Similar to UTF-8, we also provide a mechanism to recover from invalid sequences, and the recovery is optimized for accuracy
instead of other metrics. 

\section{UTF-8 based representation}
\label{sect:utf8}

UTF-8 based models have been proposed for natural language processing (NLP) \cite{gillick2016} \cite{ruiz2017byte} \cite{xue2021}.
The idea is to convert text to a sequence of variable-length UTF-8 codewords, and to have the model predict one byte at each decoding step.
The advantages of byte-level representation are compactness and universality, as any combination of languages may be represented with an output dimension of only 256.
However, a sequence represented at byte level is often longer than its character-level counterpart, especially for CJK languages~\cite{wang2020neural}.
This is because while Latin characters are represented by a single byte, many CJK characters and accented characters are represented by multiple bytes.
As a result, a byte-level model can be error-prone since it needs to make multiple predictions for many single characters,
and each prediction might make a mistake. 

To compensate for the drawback of making byte level mistakes, \cite{li2019, deng2022} propose byte-level subwords for E2E ASR.
The idea is to apply byte pair encoding (BPE) \cite{sennrich2016neural} to UTF-8 codeword sequences to create UTF-8 subwords. As subwords are in general longer than
byte-level tokens, this approach reduces the number of steps required by the decoding process.
However, BPE does not guarantee that the output will be a valid UTF-8 sequence. To repair an invalid byte sequence,
\cite{li2019} proposes a dynamic programming algorithm to recover as many characters as possible given any byte sequence.
While this dynamic programming approach ensures the output sequence is always valid, it optimizes for the number of valid characters, not ASR quality.

\section{Optimizing byte-level representation}
\label{sect:latent}

\subsection{Formulation as a latent variable optimization problem}
We first formulate the representation problem as an optimization problem with latent variables.
Suppose $W$ is a sequence of label tokens (e.g. words, subwords or characters),
$X$ is a sequence of acoustic features and $Q$ is a sequence of discrete latent variables, $[\cdots , q_i, \cdots]$, from a vocabulary ${\cal V_Q}$,
then we can write the formula of posterior probability as,
\begin{equation}
\label{eqn:posterior_orig}
P(W|X) = \sum_Q P(W | Q, X)P(Q | X) \ .
\end{equation}
Assuming $Q$ captures sufficient information to infer $W$, we can then simplify Equation~\ref{eqn:posterior_orig} to:
\begin{equation}
\label{eqn:posterior}
P(W|X) = \sum_Q P(W | Q)P(Q | X) \ .
\end{equation}
The representation problem becomes the optimization problem of finding the set of latent variables ${\cal V_Q}$,
such that the posterior probability of Equation~\ref{eqn:posterior} is maximized.
In our case, the sequence $Q$ is the byte-level representation to be optimized for ASR, and it forms the output of the ASR model. In this formulation, $P(Q|X)$ is the ASR model that
describes how acoustic features will be mapped to these latent variables.
$P(W|Q)$ plays a role similar to a lexicon, which describes how $Q$ can be used to infer $W$.
One difference is that this lexicon will be jointly optimized.
In this work, we assume that the relationship between $W$ and $Q$ is deterministic,
so we can convert label tokens to and from latent variables.

Optimizing Equation~\ref{eqn:posterior} directly is not trivial since $Q$ is hidden
and this equation marginalizes over all possible $Q$.
In practice, it means we would need to sample $Q$ from $P(Q|X)$ to
approximate the equation, which is still expensive and difficult.
At a high level, the challenge here is to solve the representation problem, $P(W|Q)$,
and the alignment problem, $P(Q|X)$, at the same time.
To circumvent this challenge, we propose to reformulate the task as an auto-encoding problem.

\subsection{Formulation as auto-encoding optimization problem}
\begin{figure}[htb]
    \centering
    \includegraphics[width=8.5cm]{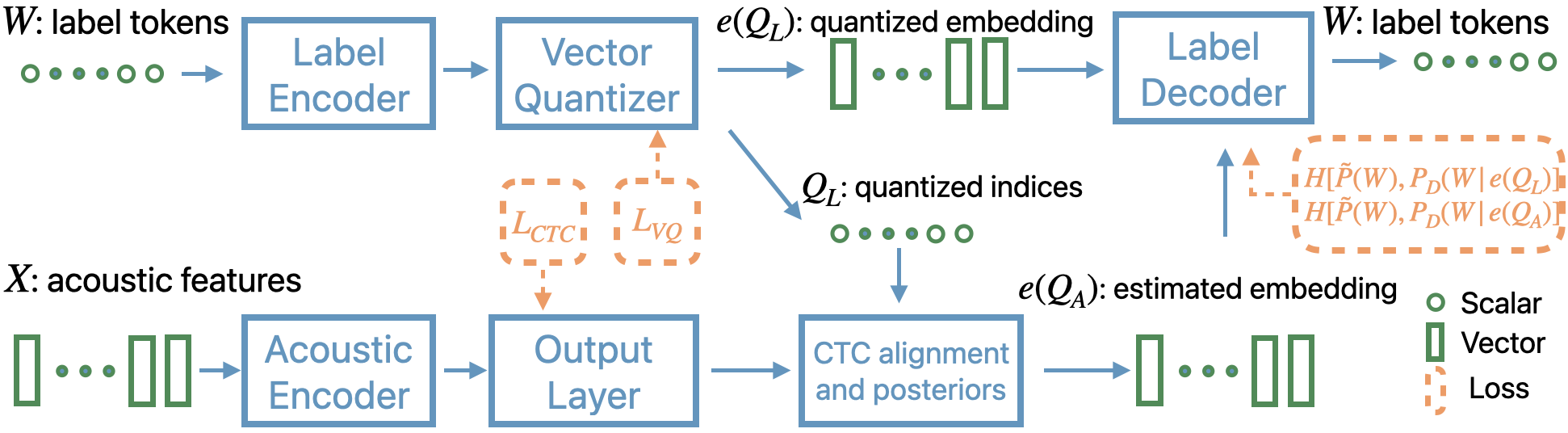}
    \caption{Overview of the proposed auto-encoder. The label tokens could be words/subwords/characters. In our experiments, the label tokens are characters.}
    \label{fig:latent_representation}
\end{figure}

The auto-encoder consists of the following components:
\begin{enumerate}
\item a label encoder, $P_L(Q | W)$;
\item an acoustic encoder, $P_A(Q | X)$;
\item a label decoder, $P_D(W | Q)$;
\item a vector quantizer, $VQ$.
\end{enumerate}
Figure~\ref{fig:latent_representation} illustrates the architecture of this auto-encoder.
This auto-encoder uses vector quantization (VQ) as its bottleneck, with the indices of the quantized embeddings serving as the latent variables.
We use $Q$ to represent the sequence of embedding indices and define $e(q)$ as a function that returns the quantized embedding vector given an embedding index $q$.
Each encoder branch represents the information that contributes to building the latent representation,
and the decoder would recover the label sequences from the latent variables. This auto-encoder can be optimized by the loss function,
\begin{eqnarray}
\label{eqn:auto_encoder}
L_{AE}& = & H[\tilde{P}(W), P_D(W | e(Q_L))] \nonumber \\
  & + & H[\tilde{P}(W), P_D(W | e(Q_A))] \nonumber \\
  & + & L_{\textit{CTC}}[P_A(Q_L|X)] \nonumber \\
  & + & L_{\textit{VQ}} \ .
\end{eqnarray}

This loss function consists of four terms. The first term is the cross entropy loss between the empirical distribution of the reference word sequence and the distribution induced by the label decoder given the output of the label encoder, $Q_L = [\cdots, q_i, \cdots ]$. This term is the standard loss of an auto-encoder, where it first uses the label encoder, $P_L(Q | W)$, to decide $Q_L$. Then, the corresponding sequence of
embeddings, $e(Q_L) = [\cdots, e(q_i), \cdots]$, is fed to the label decoder to compute $P_D(W | e(Q_L))$. This term would optimize the label decoder and the label encoder end-to-end.

The second cross entropy loss in Equation~\ref{eqn:auto_encoder} represents the loss of the decoder when it uses the information
from the acoustic encoder. $Q_A$ is a sequence of latent variables chosen by the acoustic encoder, and has the same length as $Q_L$.
However, instead of sampling from $P_A(Q|X)$ to obtain $Q_A$ explicitly, a sequence of corresponding embeddings, $[\cdots, e_i(Q_A), \cdots ]$, is obtained via alignment of the acoustic features with $Q_L$ from the label encoder, followed by a linear combination of the label embeddings,
\begin{equation}
\underbrace{[ \cdots, e_i(Q_A), \cdots]}_\text{same length as $Q_L$} = [ \cdots, \sum_{q \in {\cal V_Q}} P(q | x_{t(i)}) e(q), \cdots ] \ .
\end{equation}
Here, for each token $q_i \in Q_L$, we use the aligned feature, $x_{t(i)}$, for which $q_i$ is first emitted.
Then, we compute the posterior probabilities across all embedding vectors in the VQ codebook, $P(q | x_{t(i)})$,
and use them to construct the embedding to be fed to the label decoder, $e_i(Q_A)$.
This term would optimize the label decoder and the acoustic encoder end-to-end.

The third term is a CTC loss~\cite{graves2006} for the acoustic encoder.
This term computes how well the acoustic features are aligned to the latent variables, $Q_L$, which are decided by
the label encoder. The CTC loss would also produce the alignment information required to compute the second term. This term would optimize only
the acoustic encoder. Finally, the fourth term is the quantization loss for VQ.

\subsection{Vector quantization variational auto-encoder}

In this work, we use a vector quantization variational auto-encoder (VQ-VAE)~\cite{oord2017} as our vector quantizer. VQ-VAE discretizes real-valued vectors into some discrete
variables while still allowing a neural network to be trained end-to-end. Standard version of VQ-VAE consists of a codebook of $M$ embedding vectors. During inference,
an input vector, $z$, will be compared to all embedding vectors in the codebook. The closest one
based on euclidean distance will be chosen and VQ-VAE would output the embedding vector $e(q)$ and its index $q$.
During training, it optimizes the following loss function,
\begin{equation}
 \textit{VQLoss}(z) = || \textit{sg}[z] - e(q_z) ||_2 + \beta || z - \textit{sg}[e(q_z)] ||_2 \ ,
\end{equation}
where $z$ is the input to the VQ module; $q_z$ is the index of the closest embedding vector $e(q_z)$ to $z$; $sg$ is the stop gradient operator and $\beta$ is a tunable parameter.
This loss function aims to keep the embedding vectors close to the inputs, and this design helps the training to be more stable~\cite{oord2017}.
To enable back-propagation, it uses a straight-through estimator that copies the gradient from the downstream to the upstream directly,
as if the VQ component does not exist~\cite{oord2017}.

The representational power of VQ-VAE depends on the size of the codebook. Given a codebook of $M$ embeddings, it assumes the inputs can be grouped into $M$ clusters.
Residual VQ-VAE (RVQ-VAE) improves by applying multiple VQ-VAE iteratively to an input~\cite{lee2022}.
For RVQ-VAE, we have $N$ codebooks and each codebook has $M$ embeddings. Given an input, it is quantized by the first VQ-VAE module and the difference between the input and quantized embedding would then be quantized by the next VQ-VAE module. This process continues until it has used all the codebooks in RVQ-VAE. Theoretically, RVQ-VAE
can represent up to $M^N$ different inputs.

In this paper, we use RVQ-VAE with two or three ($N=2,3$) codebooks and the codebook size $M$ is always 256.
Therefore, each embedding index can be represented by one byte, and each label token is represented by $N$ bytes.
We choose this setting so we can compare to UTF-8 encoding, though in practice, one can pick any number of codebooks and codebook size.

\subsection{Convert bytes to labels with error correction}

When using a byte-level representation, errors made by the ASR model could create invalid byte sequences. For example, the byte sequence might contain substitution, deletion or insertion errors.
Therefore, we need an error correction mechanism, and such a mechanism should optimize for accuracy.
In our proposed VQ representation, the label decoder at inference time can perform error correction by estimating the most likely label sequence, as shown in Algorithm~\ref{alg:error_correction}.
In this algorithm, errors in a byte sequence would affect the quality of the embeddings, but the label decoder still has the potential to map those embeddings to the correct label.
Note that this algorithm is also used at training time with the multiple bytes per label token produced by RVQ-VAE, but since the byte sequence is decided by the label encoder applied to the reference text, optimization does not directly target insertion or deletion errors.
\begin{algorithm}
\caption{The bytes to labels conversion procedure for our proposed VQ based representation. The label decoder would produce the most likely labels even if the input byte sequence contains errors}\label{alg:error_correction}
\begin{algorithmic}
\Function{bytes\_to\_labels}{Q: List[int]}
    \State $W \gets []$ \Comment{the label sequence to be returned}
    \State $v \gets \overrightarrow{0}$ \Comment{a variable to store the embedding}
    \State $k \gets -1$
    \For{$i = 1, \dots, len(Q)$}
        \State $j \gets cb(Q[i])$ \Comment{get the codebook index of $Q[i]$}
        \If{$k==-1 \textit{ or } k < j$}
            \State $v \gets v + e(Q[i])$ \Comment{add the $Q[i]$'s emb to $v$}
        \Else
            \State $w \gets \argmax_w P_D(w|v)$ \Comment{label decoder}
            \State $W \gets W \cdot w$ \Comment{append $w$ to $W$}
            \State $v \gets e[Q[i]]$ \Comment{reset $v$ to $Q[i]$'s emb}
        \EndIf
        \State $k \gets j$
    \EndFor
    \State $w \gets \argmax_w P_D(w|v)$ \Comment{label decoder}
    \State $W \gets W \cdot w$ \Comment{append $w$ to $W$}
    \State \Return W
\EndFunction
\end{algorithmic}
\end{algorithm}

\subsection{Comparing UTF-8 and VQ based representation}

While UTF-8 and our proposed VQ based representations are both byte-level representations, they have a few key differences. First, UTF-8 is a variable length prefix code,
but our proposed VQ representation is a fixed length code. Second, VQ representation can be optimized for a specific machine learning task while UTF-8 is fixed and
not optimized for machine learning. Third, UTF-8 recovers invalid sequences with the sequences with most characters~\cite{li2019}, while VQ's label decoder would try to
find the most likely output given any input sequence. Finally, as a learned representation, it is possible to have collision, i.e. different characters can
be mapped to the same byte sequences, which is not the case for UTF-8.

\section{Experimental results}
\label{sect:expt}

We evaluate our approach through our proprietary English and Mandarin dictation tasks. Since our proposed approach is new, we would like
to focus on simpler cases where the ASR system only handles two languages. We build these bilingual English and Mandarin ASR models
using a mixture of monolingual English and Mandarin supervised data, which is randomly sampled and anonymized.
The English training set consists of 10k hours of data while the Mandarin training set has 14k hours of data.
Our training sets cover domains like dictation and assistant use cases. For test sets, we use the dictation test sets for English and Mandarin, which
are randomly sampled and anonymized. The English test set consists of 27 hours of data and the Mandarin test set has 13 hours of data. Both train and test sets
cover platforms like smart phones, tablets and laptops. When we report accuracy numbers on these test sets, we report word error rate (WER) for English
and character error rate (CER) for Mandarin. We use the term token error rate (TER) when we discuss the error rates of both languages.

The model we built is a CTC-AED model similar to the one described in~\cite{yao2021wenet}. The model consists of an acoustic encoder
and a cross attention decoder (AED). The input to the CTC-AED model is 80 dimensional
mel filterbank features with 25ms window and 10ms shift. A depthwise separable convolutional layer~\cite{chollet2017} would downsample the features by 6x.
The acoustic encoder consists of 12 conformer blocks~\cite{gulati20_interspeech} with eight attention heads and the hidden dimension is 512.
The attention decoder consists of six layers of bi-directional transformer blocks.
Each transformer block has eight attention heads and the hidden dimension is also 512.
The entire model has around 120M parameters. The decoding process consists of two passes.
In the first pass, we perform CTC prefix beam search with the acoustic encoder to generate N-best hypotheses. Then,
these hypotheses are rescored by the attention decoder for the final output.

We compare three types of output representations for this bilingual setup,
\begin{enumerate}
  \item Character based output
  \item UTF-8 subword output with BPE
  \item Our proposed VQ based subword output with BPE
\end{enumerate}
Both UTF-8 and VQ representation use BPE to create subword outputs. The character based representation has all English and simplified Chinese characters, and the size is around 8000.

For VQ based representation, the auto-encoder consists of three components: acoustic encoder, label encoder and label decoder.
The acoustic encoder has the same model architecture as the acoustic encoder of the CTC-AED model.
It has an output layer that predicts the embeddings used in the VQ module. This output layer is used for the CTC loss in Equation~\ref{eqn:auto_encoder}.
The label encoder consists of six uni-directional transformer blocks, and the label decoder is a linear transform that converts the input embedding to the output tokens.
The label encoder and decoder are connected by the VQ module. After we train the auto-encoder,
we use the label encoder to convert all the transcripts in our training data into byte sequences. Then, we run BPE with SentencePiece to create byte-level subwords
similar to the UTF-8 approach. These byte-level subwords will be used as the output of the CTC-AED model.

Table~\ref{tbl:utf8_vs_rvq} shows the results of the three bilingual CTC-AED models built with these output representations. Across different number of subwords
in the ASR output, VQ based representation has improvement over UTF-8 based representation. With 8000 subwords, proposed VQ based approach has 5.8\% relative reduction
in WER for English and 3.7\% relative reduction in CER for Mandarin. This suggests it is possible to optimize a byte-level representation
for a target task. Compared to character based representation, both VQ and UTF-8 are better on English but similar accuracy on Chinese.
With 8000 subwords, which has the same size as the character based output, VQ has 14.8\% and 2.3\% relative reduction in error rate for English and Mandarin respectively.
Character based output is expected to have worse English accuracy since it does not have common English subwords in the output.
In comparison, VQ and UTF-8 can have an output dimension smaller than the size of the character set. They provide more flexibility to system design.

\begin{table}[tbp]
\centering
\caption{\label{tbl:utf8_vs_rvq} TER(\%) of CTC-AED model using character based, UTF-8 and our proposed VQ based representation. The VQ based representation has three codebooks and each codebook has 256 embeddings ($N=3$, $M=256$). Both UTF-8 and VQ use BPE to create subword output with their corresponding error correction algorithms. Each column uses different number of subwords in the ASR output.}
\begin{tabular}{|c|c|c|c|c|c|}
\hline
rep. & lang & 2000 & 4000 & 8000 & 16000 \\
\hline \hline
Char & EN & - & - & 9.61 & - \\
UTF-8 & EN & 9.25 & 8.84 & 8.70 & 8.41\\
VQ & EN & 8.77 & 8.29 & 8.19 & 8.14\\
\hline
Char & ZH & - & - & 9.77 & - \\
UTF-8 & ZH & 10.53 & 10.27 & 9.92 & 9.43 \\
VQ & ZH & 10.48 & 10.06 & 9.55 & 9.14 \\
\hline
\end{tabular}
\end{table}

\subsection{Ablation study}
We would like to look at how factors, like the number of codebooks in VQ, and the four losses used in training the auto-encoder would affect the ASR accuracy.
Table~\ref{tbl:vq_num_codebooks} shows the results of comparing different number of codebooks. We are interested in the case of two codebooks since effectively,
each label token would be represented by two bytes and theoretically, it should be enough to cover 8000 English and Chinese characters. However, as shown by the
results, two codebooks are not enough and give suboptimal accuracy. It is due to the low utilization rate of the codebooks,
where many embeddings in the codebooks are inactive. This problem, known as index collapse~\cite{kaiser2018}, is thoroughly studied in~\cite{huh2023}
and we will explore those techniques in our future work.
\begin{table}[tbp]
\centering
\caption{\label{tbl:vq_num_codebooks} TER(\%) of CTC-AED model using our proposed VQ based representation with different VQ configurations. }
\begin{tabular}{|c|c|c|c|c|}
\hline
VQ config & lang & 2000 & 4000 & 8000\\
\hline \hline
$N=2, M=256$ & EN & 8.51 & 8.30 & 8.40 \\
$N=3, M=256$ & EN & 8.77 & 8.29 & 8.19 \\
\hline
$N=2, M=256$ & ZH & 11.32 & 10.96 & 10.70 \\
$N=3, M=256$ & ZH & 10.48 & 10.06 & 9.55 \\
\hline
\end{tabular}
\end{table}

Table~\ref{tbl:ac_xent_weights} shows the results of applying different weights to the cross entropy loss
with respect to the acoustic encoder (the second term of Equation~\ref{eqn:auto_encoder}). When the weight is zero, it means the acoustic encoder
would not contribute to representation learning. In this case, the auto-encoder would ignore acoustic information and the acoustic encoder would fit to the
latent variables decided by the label encoder. The results in Table~\ref{tbl:ac_xent_weights} show that acoustic information helps, and
this supports the core concept of this paper, where optimizing a representation with all available information is beneficial.
\begin{table}[tbp]
\centering
\caption{\label{tbl:ac_xent_weights} TER(\%) of CTC-AED model when the underlying VQ representations are trained with different weights to the cross entropy loss with respect to the acoustic encoder. For this experiment, the VQ module has three codebooks and each codebook has 256 embeddings ($N=3, M=256$). BPE is performed to generate 8000 byte-level subwords}
\begin{tabular}{|c|c|c|c|}
\hline
lang & 0.0 & 0.5 & 1.0 \\
\hline \hline
EN & 8.61 & 8.49 & 8.19 \\
ZH & 9.71 & 9.89 & 9.55\\
\hline
\end{tabular}
\end{table}

\section{Conclusions and future work}
\label{sect:conclusions}

In this paper, we propose a novel algorithm to optimize a byte-level representation for ASR and compare it with UTF-8 representation.
Our proposed representation can be optimized with both audio and text data, and the error correction mechanism is optimized for accuracy.
On our English/Mandarin dictation test sets, our auto-encoder and VQ based approach can achieve 5\% relative reduction in TER.

As this proposed algorithm is new, we focus on simpler bilingual ASR in this paper. To learn a representation that can cover all languages,
we believe there are multiple issues that need to be addressed, like the index collapse issue~\cite{kaiser2018}.
The index collapse issue is well studied within the machine learning community~\cite{huh2023}
and we will explore some of the ideas in that space.
We hope that as the algorithm becomes more mature, we can learn a universal byte-level representation for all languages
that functions like UTF-8.

Comparing UTF-8 and our proposed algorithm, while our approach has accuracy advantages,
UTF-8 has some desirable properties, such as, being a prefix code, not having a collision issue.
In addition, the variable length design is more flexible, though ideally, the length would be correlated with token frequency.
In the future, we will explore VQ methods that allow variable length. If code length can be optimized jointly, we would be able to bypass
BPE and use that to generate target subwords directly. While as a machine learning approach,
it cannot guarantee no collision in the learned representation, we will look into whether we can derive a prefix code from the auto-encoder.
This would also make encoding and decoding more efficient. Then, we will also look into using other side information, such as a phonemic lexicon, to improve the representation.

\section{Acknowledgment}
The authors would like to thank Pawel Swietojanski, Ossama Abdelhamid, Takaaki Hori, and  Barry Theobald for their support and useful discussions.

\bibliographystyle{IEEEbib}
\bibliography{abbrev,e2e}

\end{document}